\documentclass[english,aps,preprint]{revtex4}
\usepackage[T1]{fontenc}
\usepackage[latin1]{inputenc}
\usepackage{graphicx}
\usepackage{amssymb}
\usepackage{sublabel}

\makeatletter

\newcommand{\comment}[1]{}

\usepackage{epsfig}
\usepackage{psfig}

\usepackage{babel}
\makeatother
\begin{document}

\preprint{This line only printed with preprint option}

\title{The storage capacity of Potts models for semantic memory retrieval}

\author{Emilio Kropff}

\email{kropff@sissa.it}

\homepage{http://www.sissa.it/~ale/limbo.html}

\affiliation{SISSA, Cognitive Neuroscience \\
 via Beirut 4 \\
 34014 Trieste, Italy}

\author{Alessandro Treves}

\affiliation{SISSA, Cognitive Neuroscience \\
 via Beirut 4 \\
 34014 Trieste, Italy}

\begin{abstract}
We introduce and analyze a minimal network model of semantic memory
in the human brain\comment{ - the memory system that, as opposed to episodic
memory, stores concepts and their relationships}. The
model is a global associative memory structured as a collection of
$N$ local modules, each coding a feature, which can take $S$
possible values, with a global sparseness $a$
(the average fraction of features describing a concept).
We show that, under optimal conditions, the number $c_{M}$ of modules connected on
average to a module can range widely between very sparse connectivity
({\em high dilution}, $c_{M}/N\to 0$) and full connectivity ($c_{M}\to N$), maintaining
a global network storage capacity (the maximum number $p_c$ of
stored and retrievable concepts) that scales like $p_c\sim c_{M}S^{2}/a$, with
logarithmic corrections consistent with the constraint that each
synapse may store up to a fraction of a bit.
\end{abstract}
\maketitle

\section{introduction}

Hebbian associative plasticity appears to be the major mechanism responsible for
sculpting connections
between pyramidal neurons in the cortex, for both short- and long-range systems of synapses.
This and other lines of evidence \cite{Braitenberg_Schuz} suggest that autoassociative memory retrieval is a general
mechanism in the cortex, occurring not only at the level of local networks, but also in higher order
processes involving many cortical areas. These areas are often regarded both from the anatomical
and from the functional point of view as distinct but interacting modules, indicating that in order
to model higher order processes we must first understand better how multimodular
autoassociative memories may operate. In a class of models conceived along these
lines, neurons in local modules, interconnected through
short-range synapses, are capable of retrieving local activity patterns, which combined
across the cortex and interacting through long-range synapses, compose global
states of activity \cite{OKa+92b}. Since long-range synapses are also modified by associative plasticity,
these states can be driven by attractor dynamics, and such networks are capable of retrieving previously learned
global patterns.

This could serve as a simple model of semantic memory retrieval. The semantic memory system, as opposed to
episodic memory, stores composite concepts, e.g. objects, and their relationships.
Although information about distinct features pertaining to a given object (e.g. its
shape, smell, texture, function)
may be processed in different areas of the cortex, a cue including only
some of the features, e.g. the shape and color, may suffice to elicit retrieval of the 
entire memory representation of the object. Imaging studies show
that, though distributed across the cortex, this activity is sparse and selective, and might 
involve regions associated to the concept being retrieved, even if not directly activated 
by the cue \cite{pulvermuller_2002}. This process could
well fit a description in terms of autoassociative multimodular memory retrieval.
In this perspective, while a local module codes for diverse values of a given
feature, a combination of features gives rise to a concept, which behaves as an
attractor of the global network and is thus susceptible of retrieval. The two-level
description that characterizes this view is the principal difference with other
attempts to describe semantic memory in terms of featural representations \cite{McRae_1997}. 

In order to reduce the complexity of a full multimodular model \cite{OKa+92a,Ful+98}
one can consider a minimal model of
semantic memory, which can be thought of as a global autoassociative
memory in which the units, instead of representing, as usual,
individual neurons, represent local cortical networks retrieving one
of various ($S$) possible states of activity. The combined activity of
these units generates a global state, which follows a retrieval dynamics.
The first question arising from this proposal is how the global storage capacity
of such a network is related to the different local and global parameters.

In the following section of this paper we present the model in mathematical
terms. In the third section we compare, through a simple signal-to-noise
analysis, different model variants proposed in the literature and
extract the minimum requirements for a network of this kind to perform
efficiently in terms of storage capacity. In the fourth section we
analyze with more sophisticated techniques the simplest model endowed
with a large capacity (the sparse Potts model) and, in particular, interesting
cases such as the very sparse and the high-$S$ limits. Following
this we study modifications to the model that make it more realistic
in terms of connectivity. Finally, we relate the results from the
previous sections to a simple information capacity analysis.

\section{$S$-state fully connected networks}

Autoassociative memories are networks of $N$ units connected to one
another by weighted synapses. These synapses are trained in such a
way that the network presents, in the ideal case, a number $p$ of
preassigned attractor states, also called stored patterns, or memories,
represented by the vectors $\vec{\xi}^{\mu}$, with $\mu=1...p$.
If the state of the network is forced into the vicinity of an attractor
(e.g., by presenting a cue correlated with one of the stored patterns)
the natural dynamics of the network converges toward the attractor,
in state space, and the memory item is said to be retrieved.
A substantial amount of the literature on attractor
networks is devoted to study the relationship between the number and
type of stored patterns and the quality of retrieval.

The state of a network at a given moment is given by the state of
each of its units, $\sigma_{i}$ for $i=1...N$. The first
quantitative analyses of autoassociative memories were of binary
models \cite{Amit_1989}, in which units could reach two possible
states, $+1$ (active unit) and $-1$ (inactive unit), resembling
Ising $\frac{1}{2}$ spins. In our case, in which units do not
represent single neurons but rather local networks, we want active
units to be able to reach one of $S$ possible states, while inactive
units remain in a 'zero' state. We thus choose the notation
$\sigma_{i}=k$ for an active unit in state $k$ and $\sigma_{i}=0$
for an inactive unit. This particular choice has no effect on the
results, since all quantities can be transformed to some other
notation. On the other hand, the stored patterns $\vec{\xi}^{\mu}$
can be simply thought of as special states of the network. For this
reason, it is natural to choose the same kind of representation for
the activity of a unit $i$ in pattern $\mu$, $\xi_{i}^{\mu}$.

Although in the first binary models of autoassociative memories
patterns where constructed with a distribution of equally probable
active and inactive units, the search of an accurate description of
activity in the brain made it necessary to introduce sparse
representations. This property of autoassociative memories is
described by the sparseness $a$, defined as the average activity (the average
fraction of active
units) in the stored patterns. In our case, because we are assuming all
$S$ different activity states to be equally probable, we consider
patterns defined by the following probability distribution
\begin{eqnarray}
P(\xi_{i}^{\mu}=0)&=&1-a\nonumber\\
P(\xi_{i}^{\mu}=k)&=&\tilde{a}\equiv\frac{a}{S}\label{eq:distribution}
\end{eqnarray}
for any active state $k$. In this way the probability to find an
active unit in a pattern is the sparseness $a$. For {\em sparse} codes,
this quantity is closer to $0$ than to $1$.

Following the assumption of Hebbian learning and, as is usual for
a simplified analysis, symmetry in the weights ($J_{ij}=J_{ji}$), a general form for
the weights is
\begin{equation}
J_{ij}^{kl}=\frac{1}{E}\sum_{\mu=1}^{p}v_{\xi_{i}^{\mu}k}v_{\xi_{j}^{\mu}l}\label{eq:weights}\end{equation}
where $E$ is some normalization constant and $v_{mn}$ is an operator
computing interactions between two states.

As one can notice, the long-range synapse weights in Eq.
\ref{eq:weights} have different values for different pre- and post-
synaptic states $k$ and $l$. In this way we do not intend to model
the actual distribution of synapses going from one cortical area to
another (since they connect neurons and not abstract states), but
rather the general mechanism of communication between these areas.
In a recent study \cite{Mechelli_2003}, the authors have raised
the issue of finding the most suitable description of global
cortical networks in terms of single long-range synapses connecting
distant local areas. Applying statistical tools
(Dynamic Causal Modeling), they propose that MRI data can be described
as produced by networks with category specific forward connections, roughly the kind of
connections modelled by Eq. \ref{eq:weights}.

The state of generic unit $i$ is determined by its local fields $h_{i}^{k}$,
which sum the influences by other units in the network and are defined as
\begin{equation}
h_{i}^{k}=\sum_{j\ne i}\sum_{l}J_{ij}^{kl}u_{\sigma_{j}l}-U(1-\delta_{k0})\label{eq:field}\end{equation}
where we introduce the operators $u_{mn}$, analogous to $v_{mn}$,
and a second (threshold) term, which has the function of regulating
the activity level across the network \cite{Buhmann_1989,Tsodyks_1988}.
The unit $i$ updates its state $\sigma_{i}$, with an asynchronous dynamics, in
order to maximize the local field $h_{i}^{\sigma_{i}}$. In the general case,
the probability to choose the state $k$ is defined as \[
P(\sigma_{i}=k)=\frac{\exp(\beta h_{i}^{k})}{\sum_{l=0}^{S}\exp(\beta h_{i}^{l})}\]
where $\beta$ is a parameter analogous to an inverse temperature.

Finally, we can include all of these elements, as is usual for the
study of attractor networks, into a Hamiltonian framework. The Hamiltonian
representation of binary networks can be extended to $S$-state
models as
\begin{equation}
H=-\frac{1}{2}\sum_{i,j\ne i}^{N}\sum_{k,l}^{S}J_{ij}^{kl}u_{\sigma_{i}k}u_{\sigma_{j}l}+U\sum_{i}^{N}\sum_{k\ne0}^{S}u_{\sigma_{i}k}\label{eq:ham}\end{equation}
Note that for the case $S=1$, Eq. \ref{eq:ham} generalizes the Hamiltonians
used in binary networks, given appropriate definitions of the weights
$J_{ij}^{kl}$ and of the operators $u_{mn}$.

We now specify a form for the $u_{mn}$ and $v_{mn}$ operators. In
the simplest and most symmetric case these operators have two
alternative values, depending on whether $m$ and $n$ are equal or
different states
\begin{eqnarray}
u_{mn}&=&(\kappa_{u}\delta_{mn}+\lambda_{u})\nonumber\\
v_{mn}&=&(\kappa_{v}\delta_{mn}+\lambda_{v})(1-\delta_{n0})\label{eq:operators}
\end{eqnarray}
where we have introduced four parameters. Particular choices for
these parameters define the different models in which we are interested,
including several proposed in the literature. In the $v$ operators,
which define the value of the weights, we have included a factor which
ensures $J_{ij}^{kl}=0$ if either $k$ or $l$ are the zero state,
to implement the idea that Hebbian learning occurs only with active
states. As we will see below, this appears to be a crucial
element in the model.

\section{Signal-to-noise analyses}

We now show that, within the group of models defined in the previous
section, there is a family (which we call 'well behaved') that
exploit multiple states and sparseness in an optimal way in terms of storage
capacity or, as usual, of $\alpha\equiv p/N$. We begin
by applying an adjusted version of the arguments developed in \cite{Buhmann_1989}.

A signal-to-noise analysis is a simplified way to estimate the stability
of stored patterns by studying what happens to a generic unit $i$ during the perfect retrieval
of a given pattern, assessing whether the state of this unit is likely to be
stable or not. We can choose this retrieved pattern to be $\vec{\xi}^{1}$
without loss of generality. Eq. \ref{eq:field} can then be rewritten
as\begin{equation}
h_{i}^{k}=\frac{1}{E}v_{\xi_{i}^{1}k}\sum_{j\neq i}\sum_{l}u_{\sigma_{j}l}v_{\xi_{j}^{1}l}+\frac{1}{E}\sum_{\mu>1}v_{\xi_{i}^{\mu}k}\sum_{j\neq i}\sum_{l}u_{\sigma_{j}l}v_{\xi_{j}^{\mu}l}-U(1-\delta_{k0})\label{eq:sig-noise}\end{equation}
where the terms in the RHS stand for signal ($\varsigma$), noise
($\rho$) and threshold respectively. Generally speaking, if the field had only the
signal part then the state would be stable, but the noise can destabilize
it.

As usual in this kind of analysis, we consider the contribution
of the noise term in Eq. \ref{eq:sig-noise} as if it were a normally
distributed random variable, i.e. through its average and its
standard deviation. In general both quantities scale like $p$, but
in some special cases the average noise is zero and the standard
deviation scales only like $\sqrt{p}$, which means that one can
store more patterns, as the noise level is reduced. It is clear that the well
behaved family of models which we are looking for must fit into this
favorable situation. As we said, a necessary but not sufficient
condition is the average of the noise to be zero. There are two ways
of imposing this into the model. The first way is to make
$\lambda_{u}=-\tilde{a}\kappa_{u}$, but in this case the standard
deviation still scales like $p$. The second way is to use
\begin{equation}
\lambda_{v}=-\tilde{a}\kappa_{v}\label{eq:condition}\end{equation}
 which makes the standard deviation scale like $\sqrt{p}$. Including
this condition, the average signal and the standard deviation of the
noise are
\begin{eqnarray}
\varsigma&=&\frac{N\kappa_{v}^{2}}{E}\kappa_{u}\tilde{a}(1-\tilde{a})S(\delta_{\xi_{i}^{1}k}-\tilde{a})(1-\delta_{k0})\nonumber\\
\rho&=&\frac{N\kappa_{v}^{2}}{E}\kappa_{u}\tilde{a}(1-\tilde{a})\sqrt{\alpha a\left\{ 1-\tilde{a}\left[1-\left(1-\frac{\lambda_{u}}{\tilde{a}\kappa_{u}}\right)^{2}\right]\left[\frac{1-a}{1-\tilde{a}}\right]\right\} }(1-\delta_{k0})\nonumber
\end{eqnarray}
where terms of order $1/N$ have been discarded.

The storage capacity $\alpha_{c}$ can be estimated as the largest
value of $\alpha$ for which $h_{i}^{\xi_{i}^{1}}$ is still likely to be the largest
among all $S+1$ local fields. The situation is quite different
depending on whether $\xi_{i}^{1}$ is in an active state or not, so one needs
to analyze both cases.
Note first that $h_{i}^{0}=0$, so if
$\xi_{i}^{1}=0$ the rest of the local fields must be negative. For
this to hold true at least within one standard deviation of the noise distribution
we require $\varsigma-U\pm\rho<0$, or in other words\[ a+\frac{U\:
E}{N\kappa_{v}^{2}\kappa_{u}\tilde{a}(1-\tilde{a})}>\sqrt{\alpha
a\left\{
1-\tilde{a}\left[1-\left(1-\frac{\lambda_{u}}{\tilde{a}\kappa_{u}}\right)^{2}\right]\left[\frac{1-a}{1-\tilde{a}}\right]\right\}
}\] where we have adopted a positive $\kappa_{u}$.

In the case in which $\xi_{i}^{1}$ is not the zero state two conditions must
be fulfilled, namely $h_{i}^{\xi_{i}^{1}}>h_{i}^{0}$ and
$h_{i}^{\xi_{i}^{1}}>h_{i}^{k\ne\xi_{i}^{1}}$. These conditions can
be condensed into\[ S(1-\tilde{a})-\frac{U\:
E}{N\kappa_{v}^{2}\kappa_{u}\tilde{a}(1-\tilde{a})}>\sqrt{\alpha
a\left\{
1-\tilde{a}\left[1-\left(1-\frac{\lambda_{u}}{\tilde{a}\kappa_{u}}\right)^{2}\right]\left[\frac{1-a}{1-\tilde{a}}\right]\right\}
}\]

The most stringent of these 2 conditions determines $\alpha_{c}$.
By choosing a suitable threshold
$U=\frac{N}{E}\kappa_{v}^{2}\kappa_{u}\tilde{a}(1-\tilde{a})\left[\frac{S}{2}-a\right]$
both conditions are made equivalent, thus optimizing the storage
capacity. This choice determines a storage capacity
of\begin{equation} \alpha_{c}\simeq\frac{S^{2}}{4a}\left\{
1-\tilde{a}\left[1-\left(1-\frac{\lambda_{u}}{\tilde{a}\kappa_{u}}\right)^{2}\right]\left[\frac{1-a}{1-\tilde{a}}\right]\right\}
^{-1}\label{eq:storage}\end{equation}

Note that the expression between curly brackets is equal to or greater
than $1-\tilde{a}$. As a consequence, the system remains optimal
as long as this expression remains of order $1$, which, considering always $a$ to be
closer to $0$ than to $1$, occurs when the expression $\left(1-\frac{\lambda_{u}}{\tilde{a}\kappa_{u}}\right)^{2}$
remains of order $1$. For this to be true we must impose \begin{equation}
\left|\lambda_{u}\right|\lesssim\tilde{a}\kappa_{u}\label{eq:condition2}\end{equation}

We thus define the well behaved models as those which fulfil the conditions
given by Eq. \ref{eq:condition} and Eq. \ref{eq:condition2}. This
simple analysis indicates that the storage capacity of models in the
well behaved family scales like $S^{2}/a$.

In the following subsections we examine different models proposed
in literature, both within and outside the well behaved family.

\subsection{Symmetric Potts model}

The symmetric Potts model was the first $S$-state neural network to be
proposed \cite{potts-kanter}. Its units can reach $S$ equivalent
states but no zero state. Though simple, a model constructed with
these elements is enough to show the $S^{2}$ behavior of the storage
capacity, as we will see. It is defined by setting\[ a=1\]
 \[
U=0\]
 two conditions related to each other (if there is no zero state,
the selectivity mechanism provided by the threshold is not necessary).
Moreover $E=S^{2}N$, which is just a normalization, and
\begin{eqnarray}
\kappa_{u}&=&\kappa_{v}=S\nonumber\\
\lambda_{u}&=&\lambda_{v}=-1\nonumber
\end{eqnarray}
 The conditions given by Eq. \ref{eq:condition} and Eq. \ref{eq:condition2}
are fulfilled, and the storage capacity in Eq. \ref{eq:storage} is
approximately\[
\alpha_{c}\approx\frac{S^{2}}{4}\]
 provided $S$ is large enough. The symmetric Potts model is then
a well behaved model of sparseness $a=1$.

This model is studied analytically with replica tools in \cite{potts-kanter},
where the author finds an $S(S-1)$ behavior of the storage capacity
for low values of $S$. Unfortunately, the cited work lacks an analysis
for high values of $S$, which is the interesting limit for modeling
multi-modular networks. It is not too difficult, however, to clarify
the behavior in this limit.

The replica storage capacity is defined as the highest value of $\alpha$
for which there is a solution to the equation

\begin{equation}
y=\frac{-1+S\int Dz[\phi(z+y)]^{S-1}}{\sqrt{\frac{\alpha(S-1)}{S}}+\int zDz\{[\phi(z+y)]^{S-1}+(S-1)\phi(z-y)[\phi(z)]^{S-2}\}}\label{eq:potts}\end{equation}
where \begin{equation}
\phi(z)\equiv\frac{1+{\rm erf}(\frac{z}{\sqrt{2}})}{2}\label{eq:fi}\end{equation}
 Throughout this paper we use the gaussian differential $Dz\equiv\frac{e^{-\frac{z^{2}}{2}}}{\sqrt{2\pi}}dz$,
and the integration limits, if not specified, are -$\infty$ and $\infty$.

We note that in Eq. \ref{eq:potts} expressions of the form $\left[\phi(z)\right]^{S}$
can be approximated by displaced Heaviside functions for high values
of $S$. Using this we obtain an approximated analytical expression
for the storage capacity

\begin{equation}
\alpha_{c}=\left[\frac{\phi(\sqrt{\frac{\pi}{2}})}{\sqrt{\frac{\pi}{2}}+\sqrt{2}
{\rm erf}^{-1}(1-\frac{\ln(2)}{S})}\right]^{2}S^{2}\label{eq:highq}\end{equation}

The factor between brackets in this equation behaves like
$\ln(S)^{-\frac{1}{2}}$ for high values of $S$, which means that the
correction for high $S$ to Kanter's low $S$ approximation is a
factor of order $\ln(S)^{-1}$.

\begin{figure}[h]
\centerline{\hbox{\epsfig{figure=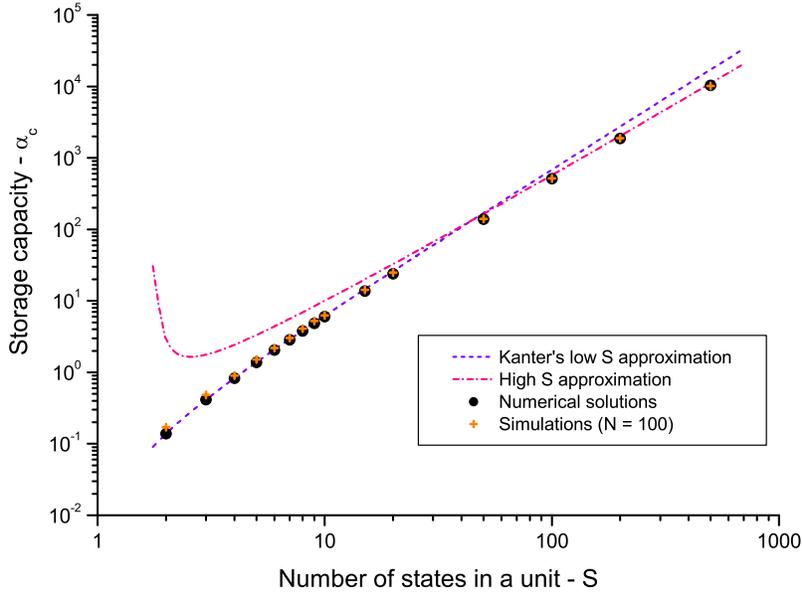,width=12cm,angle=0}}}
\caption{Storage capacity of a symmetric Potts network of $N = 100$
units for increasing $S$. Both axes are logarithmic. Black dots show
numerical solutions for Eq. \ref{eq:potts}, which overlap almost
perfectly with the simulations (plus signs). For low values of $S$
($S\lesssim50$) Kanter's low $S$ approximation fits well, while the
high values of $S$ are well fitted by Eq.
\ref{eq:highq}.}\label{fig:potts capacity}
\end{figure}

We show in Fig. \ref{fig:potts capacity} the results of simulations
of a symmetric Potts network ($N=100$) contrasted with Kanter's low
$S$ approximation and our own high $S$ approximation of Eq.
\ref{eq:highq}. The analytical predictions fit tightly the results
of the simulations, both for low and high $S$.

\subsection{Biased Potts model}

This model is proposed and studied in \cite{Bolle_1993}. The authors
extend the symmetric Potts model to an $S$-state network with arbitrary
probability distribution for the states of the units in stored patterns.
We adapt their formalism to the case of $S$ equivalent states, a
zero state and sparseness $a$. The parameters are then

\begin{equation}
\begin{array}{lll}
U&=&0\\
E&=&N\\
u_{mn}&=&((S+1)\delta_{mn}-1)\\
v_{mn}&=&(\delta_{mn}-P_{n})
\end{array}\label{eq:wrong}
\end{equation}
where $P_{k}$ is the probability of a unit in the stored patterns
to be in state $k$. This model does not fit exactly our description
because the $v$ operators generate weights $J_{ij}^{kl}$ that are
not necessarily zero when $k$ or $l$ are zero. The signal to noise analysis
for this situation shows a very poor storage capacity, scaling like
$a^{2}$. If one adds a non-zero threshold ($U\sim a\: S$ in the
optimal case) the storage capacity grows but remains of order $1$.
These two results show that allowing for non-zero weights
to connect zero states is a drawback for the system. The
poor performance can, however, be improved by multiplying the $v$
operators by the corresponding $(1-\delta_{n0})$ factors, and by
adding a threshold. In this way, instead of Eq. \ref{eq:wrong} we
introduce our definition, Eq. \ref{eq:operators}, for the $v$
operators, with the values for $\kappa$'s and $\lambda$'s arising
naturally from the model as
\begin{eqnarray}
\kappa_{u}&=&S+1\nonumber\\
\lambda_{u}&=&-1\nonumber\\
\kappa_{v}&=&1\nonumber\\
\lambda_{v}&=&-\tilde{a}\nonumber\\
U&\sim &aS\nonumber
\end{eqnarray}

As in the symmetric Potts model, the condition given by Eq. \ref{eq:condition}
is fulfilled. However, the second condition (Eq. \ref{eq:condition2})
can be approximated for high $S$ by \comment{$\[a\moresim 1/(1+1/s)\sim 1\]$}\[a\gtrsim 1/(1+1/S)\sim 1\] 
which does not stand true for sparse coding.
If, instead, $a\ll1$, the critical value of $\alpha$
in Eq. \ref{eq:storage} can be approximated as\[
\alpha_{c}\approx\frac{S^{2}}{4a}\left\{ 1+\frac{1}{a\: S}\right\} ^{-1}\]
Hence the storage capacity of the biased Potts model can be preserved
close to optimal by imposing an {\em ad hoc} relation between two parameters
that are a priori independent, to assure $1\ll a\: S$. In this particular situation the model is well behaved. In the opposite limit, when $a\: S\ll1$, the storage
capacity scales like $S^{3}$, which is inferior to the
$S^{2}/a$ behavior of the well behaved family.

\subsection{Sparse Potts model.}

The simplest version of a well behaved model is perhaps the one introduced as a
model for semantic memory \cite{Treves_2005}, with the parameter values
\begin{eqnarray}
E&=&Na(1-\tilde{a})\nonumber\\
\kappa_{u}&=&\kappa_{v}=1\nonumber\\
\lambda_{u}&=&0\nonumber\\
\lambda_{v}&=&-\tilde{a}\nonumber\\
U&\sim&1/2\nonumber
\end{eqnarray}
With these parameters, the sparse Potts model is clearly well behaved,
and the storage capacity in Eq. \ref{eq:storage} becomes

\[
\alpha_{c}\simeq \frac{S^{2}}{4a}\]

\section{replica analysis}

Having introduced a simple model with optimal storage capacity, we
can proceed to analyze the corrections to the signal-to-noise
estimation by treating the problem in a more refined way with the
classical replica method. The Hamiltonian in Eq. \ref{eq:ham} can be
rewritten for the sparse Potts model as \[
H=-\frac{1}{2}\sum_{i,j\ne i}^{N}\sum_{k,l}^{S}J_{ij}^{kl}\delta_{\sigma_{i}k}\delta_{\sigma_{j}l}+U\sum_{i}^{N}(1-\delta_{\sigma_{i}0})\]
 with\[
J_{ij}^{kl}=\frac{1}{Na(1-\tilde{a})}\sum_{\mu=1}^{p}(\delta_{\xi_{i}^{\mu}k}-\tilde{a})(\delta_{\xi_{j}^{\mu}l}-\tilde{a})(1-\delta_{k0})(1-\delta_{l0})\]
 constructed using\[
v_{mn}=(\delta_{mn}-\tilde{a})(1-\delta_{n0})\]

We consider the limit $p\rightarrow\infty$ and $N\rightarrow\infty$
with the ratio $\alpha\equiv\frac{p}{N}$ fixed. Patterns with index
$\nu$ ($\mu)$ are condensed (not condensed). Following the replica
analysis \cite{Amit_1989} the free energy can be calculated as
\begin{eqnarray}
\lefteqn{f={\scriptstyle \begin{array}{c}\\lim\\^{n\rightarrow0}\end{array}}\frac{a(1-\tilde{a})}{2n}\sum_{\rho=1}^{n}\sum_{\nu}(m_{\rho}^{\nu})^{2}+} \nonumber\\
  & & +\frac{\alpha}{2n\beta}Tr\left(\ln[a(1-\tilde{a})(\mathbb{I}-\beta\tilde{a}\mathbf{q})]\right)+\frac{\alpha\beta\tilde{a}^{2}}{2n}\sum_{\rho,\lambda=1}^{n}q_{\rho\lambda}r_{\rho\lambda}+\frac{\tilde{a}}{n}(\frac{\alpha}{2}+U\: S)\sum_{\rho=1}^{n}q_{\rho\rho}- \nonumber\\
  & & -\frac{1}{n\beta}\left\langle \left\langle \ln Tr_{\sigma_{\rho}}\exp\left\{ \beta\sum_{\rho=1}^{n}\sum_{\nu}m_{\rho}^{\nu}v_{\xi^{\nu}\sigma_{\rho}}+\frac{\alpha\beta^{2}}{2S(1-\tilde{a})}\sum_{\rho,\lambda=1}^{n}r_{\rho\lambda}\sum_{k}P_{k}v_{k\sigma_{\rho}}v_{k\sigma_{\lambda}}\right\} \right\rangle \right\rangle\nonumber
  \end{eqnarray}
where $P_{k}$ is the probability of a neuron to be in state $k$
in a stored pattern, as defined in Eq. \ref{eq:distribution}. The order parameters $m$ stand for the overlaps of the 
states with
different patterns, and $q_{\rho\lambda}$ is analogous to the
Edward-Anderson parameter \cite{Edwards-Anderson}, with the
following definitions
\begin{eqnarray}
m_{\rho}^{\nu}&=&\frac{1}{N\:a(1-\tilde{a})}\left\langle \left\langle \sum_{i=1}^{N}\left\langle
v_{\xi_{i}^{\nu}\sigma_{i}^{\rho}}\right\rangle \right\rangle\right\rangle \nonumber\\
q_{\rho\lambda}&=&\frac{1}{N\:\tilde{a}\: a(1-\tilde{a})}\sum_{i=1}^{N}\left\langle \left\langle \sum_{k}P_{k}\left\langle v_{k\sigma_{i}^{\rho}}v_{k\sigma_{i}^{\lambda}}\right\rangle \right\rangle \right\rangle \nonumber\\
r_{\rho\lambda}&=&\frac{S(1-\tilde{a})}{\alpha}\sum_{\mu}\left\langle \left\langle m_{\rho}^{\mu}m_{\lambda}^{\mu}\right\rangle \right\rangle -\left(\frac{2S\: U}{\alpha}+1\right)\frac{\delta_{\rho\lambda}}{\beta\tilde{a}} \nonumber
\end{eqnarray}
in such a way that they are all of order $1$. Consider, for example,
that if $\sigma_{i}^{\rho}=\xi_{i}^{\nu}$ for all $i$ then $m_{\rho}^{\nu}=1$
on average, while $m_{\rho}^{\nu}=0$ on average if both quantities
are independent variables.

We now make two assumptions. First, we consider for simplicity that
there is only one condensed pattern, making the index $\nu$ superflous.
Second, we assume that there is replica symmetry, and substitute
\begin{eqnarray}
m_{\rho}^{\nu}&=&m\nonumber\\
q_{\rho\lambda}&=&\left\{ \begin{array}{cc}
q & if\:\rho\neq\lambda\nonumber\\
\tilde{q} & if\:\rho=\lambda\nonumber\end{array}\right.\nonumber\\
r_{\rho\lambda}&=&\left\{ \begin{array}{cc}
r & if\:\rho\neq\lambda\nonumber\\
\tilde{r} & if\:\rho=\lambda\nonumber\end{array}\right.\nonumber
\end{eqnarray}
Taking this into account, we arrive to the final expression for the
free energy
\begin{eqnarray}
\lefteqn{f=a(1-\tilde{a})\frac{m^{2}}{2}+\frac{\alpha}{2\beta}\left[\ln\left(a(1-\tilde{a})\right)+\ln(1-\tilde{a}C)-\frac{\beta q\tilde{a}}{(1-\tilde{a}C)}\right]+}\nonumber\\
 & & +\frac{\beta\alpha\tilde{a}^{2}}{2}(\tilde{q}\tilde{r}-qr)+\left[\frac{\alpha}{2}+S\: U\right]\tilde{q}\tilde{a}-\frac{1}{\beta}\left\langle \left\langle \int D\mathbf{z}\ln\left(1+\sum_{\sigma\neq0}\exp(\beta\mathcal{H}_{\sigma}^{\xi})\right)\right\rangle \right\rangle \nonumber
\end{eqnarray}
 where the finite-valued variable $C$ has been introduced\[
C\equiv\beta(\tilde{q}-q)\]
 in such a way that it is of order $1$ and\begin{equation}
\mathcal{H}_{\sigma}^{\xi}\equiv m\: v_{\xi\sigma}-\frac{\alpha a}{S^{2}}\frac{\beta(r-\tilde{r})}{2}(1-\delta_{\sigma0})+\sum_{k}\sqrt{\frac{\alpha r\: P_{k}}{S(1-\tilde{a})}}z_{k}v_{k\sigma}\label{eq:HH}\end{equation}
 Note that $\mathcal{H}_{0}^{\xi}=0$.

We now derive the fixed-point equation for $m$ as an example of how
the limit $\beta\rightarrow\infty$ is taken. The equation for finite
$\beta$ is\[
m=\frac{1}{a(1-\tilde{a})}\left\langle \left\langle \int D\mathbf{z}\sum_{\sigma}v_{\xi\sigma}\left[\frac{1}{1+\sum_{\rho\neq\sigma}\exp\left\{ \beta(\mathcal{H}_{\rho}^{\xi}-\mathcal{H}_{\sigma}^{\xi})\right\} }\right]\right\rangle \right\rangle \]

In the limit $\beta\rightarrow\infty$ the expression between brackets
is $1$ if $\mathcal{H}_{\sigma}^{\xi}>\mathcal{H}_{\rho}^{\xi}$
for every $\rho\ne\sigma$ and $0$ otherwise. It can be thus expressed
as a product of Heaviside functions. The equation for $m$ at zero
temperature is then\[
m=\frac{1}{a(1-\tilde{a})}\sum_{\sigma\neq0}\left\langle \left\langle \int D\mathbf{z}\: v_{\xi\sigma}\prod_{\rho\neq\sigma}\Theta\left[\mathcal{H}_{\sigma}^{\xi}-\mathcal{H}_{\rho}^{\xi}\right]\right\rangle \right\rangle \]
 In the same way we derive the rest of the fixed point equations at
zero temperature \comment{
			q{\scriptstyle \begin{array}{c}
			\\{\textstyle {\displaystyle \longrightarrow}}\\
			^{\beta\rightarrow\infty}\end{array}}\tilde{q}=\frac{1}{a}\sum_{\sigma\neq0}\left\langle \left\langle \int D\mathbf{z}\prod_{\rho\neq\sigma}\Theta\left[\mathcal{H}_{\sigma}^{\xi}-\mathcal{H}_{\rho}^{\xi}\right]\right\rangle \right\rangle \]
			 \[
			C=\frac{1}{\tilde{a}^{2}\sqrt{\alpha r}}\sum_{\sigma\neq0}\sum_{k}\left\langle \left\langle \int D\mathbf{z}\sqrt{\frac{P_{k}}{S(1-\tilde{a})}}v_{k\sigma}z_{k}\prod_{\rho\neq\sigma}\Theta\left[\mathcal{H}_{\sigma}^{\xi}-\mathcal{H}_{\rho}^{\xi}\right]\right\rangle \right\rangle \]
			 \[
			\tilde{r}{\scriptstyle \begin{array}{c}
			\\{\textstyle {\displaystyle \longrightarrow}}\\
			^{\beta\rightarrow\infty}\end{array}}r=\frac{q}{(1-\tilde{a}C)^{2}}\]
			\begin{equation}
			\beta(r-\tilde{r})=2U\frac{S^{2}}{a\alpha}-\frac{C}{1-\tilde{a}C}\label{eq:multiple fix}\end{equation}}
\begin{equation}
\begin{array}{c}
q{\scriptstyle \begin{array}{c}
\\{\textstyle {\displaystyle \longrightarrow}}\\
^{\beta\rightarrow\infty}\end{array}}\tilde{q} = \frac{1}{a}\sum_{\sigma\neq0}\left\langle \left\langle \int D\mathbf{z}\prod_{\rho\neq\sigma}\Theta\left[\mathcal{H}_{\sigma}^{\xi}-\mathcal{H}_{\rho}^{\xi}\right]\right\rangle \right\rangle \\
C=\frac{1}{\tilde{a}^{2}\sqrt{\alpha r}}\sum_{\sigma\neq0}\sum_{k}\left\langle \left\langle \int D\mathbf{z}\sqrt{\frac{P_{k}}{S(1-\tilde{a})}}v_{k\sigma}z_{k}\prod_{\rho\neq\sigma}\Theta\left[\mathcal{H}_{\sigma}^{\xi}-\mathcal{H}_{\rho}^{\xi}\right]\right\rangle \right\rangle\\
\tilde{r}{\scriptstyle \begin{array}{c}
\\{\textstyle {\displaystyle \longrightarrow}}\\
^{\beta\rightarrow\infty}\end{array}}r=\frac{q}{(1-\tilde{a}C)^{2}}\\
\beta(r-\tilde{r})=2U\frac{S^{2}}{a\alpha}-\frac{C}{1-\tilde{a}C}
\end{array}\label{eq:multiple fix}
\end{equation}
The differences between $r$ and $\tilde{r}$, and between $q$ and
$\tilde{q}$, are of order $\frac{1}{\beta}$. From the last equation
it can be seen that the threshold $U$ has the effect of changing the
sign of $(r-\tilde{r})$ and allowing $\alpha$ to scale like
$\frac{S^{2}}{a}$, with the variables $C$, $r$ and $\tilde{r}$, as
we have said, of order $1$ with respect to $a$ and $S$.

\subsection{Reduced saddle-point equations}

It is possible to calculate the averages in Eqs. \ref{eq:multiple
fix} by reducing the problem to the following variables, which represent
respectively signal and noise contributions\[
y\equiv
m\sqrt{\frac{S^{2}}{\alpha a}\frac{(1-\tilde{a})}{r}}\equiv
m\sqrt{\frac{(1-\tilde{a})}{\tilde{\alpha}r}}\]
 \[
x\equiv\frac{\tilde{\alpha}\beta(r-\tilde{r})}{2}\sqrt{\frac{(1-\tilde{a})}{\tilde{\alpha}r}}\]
where we have introduced the normalized (order $1$) storage capacity
$\tilde{\alpha}\equiv\alpha a/S^{2}$, which clarifies that both variables
$x$ and $y$ are also of order $1$.

At the saddle point, using equations \ref{eq:multiple fix}, we obtain
\begin{equation}
\begin{array}{lll}
y&=&\sqrt{\frac{1-\tilde{a}}{\tilde{\alpha}}}\left(\frac{m}{\sqrt{q}+C\sqrt{r}}\right)\\
x&=&\sqrt{\frac{1-\tilde{a}}{\tilde{\alpha}}}\left[U-\tilde{\alpha}C\sqrt{\frac{r}{q}}\right]\left[\frac{1}{\sqrt{q}+\tilde{a}C\sqrt{r}}\right]
\end{array}\label{eq:xy}
\end{equation}
 which shows that the relevant quantities to describe the system are
$m,$ $q$, and $C\sqrt{r}$. Following this we compute the averages
and get from Eq. \ref{eq:multiple fix} the corresponding equations
in terms of $y$ and $x$
\begin{eqnarray}
\lefteqn{q=\frac{(1-a)}{\tilde{a}}\int Dw\int_{y\tilde{a}+x-i\sqrt{\tilde{a}}w}^{\infty}Dz\phi(z)^{S}+}\nonumber\\
& &+\int Dw\int_{-y(1-\tilde{a})+x-i\sqrt{\tilde{a}}w}^{\infty}Dz\phi(z+y)^{S}+(S-1)\int Dw\int_{y\tilde{a}+x-i\sqrt{\tilde{a}}w}^{\infty}Dz\phi(z-y)\phi(z)^{S-1}\nonumber\\
\lefteqn{m=\frac{1}{1-\tilde{a}}\int Dw\int_{-y(1-\tilde{a})+x-i\sqrt{\tilde{a}}w}^{\infty}Dz\phi(z+y)^{S}-q\frac{\tilde{a}}{1-\tilde{a}}}\nonumber\\
\lefteqn{C\sqrt{r}=\frac{1}{\sqrt{\tilde{\alpha}(1-\tilde{a})}}\left\{ \frac{(1-a)}{\tilde{a}}\int Dw\int_{y\tilde{a}+x-i\sqrt{\tilde{a}}w}^{\infty}Dz(z+i\sqrt{\tilde{a}}w)\phi(z)^{S}+\right.}\nonumber\\
& & +\int Dw\int_{-y(1-\tilde{a})+x-i\sqrt{\tilde{a}}w}^{\infty}Dz(z+i\sqrt{\tilde{a}}w)\phi(z+y)^{S}+\nonumber\\
& & +\left.(S-1)\int Dw\int_{y\tilde{a}+x-i\sqrt{\tilde{a}}w}^{\infty}Dz(z+i\sqrt{\tilde{a}}w)\phi(z-y)\phi(z)^{S-1}\right\}\label{eq:multiple 2}
\end{eqnarray}

\begin{figure}[h]
\centerline{\hbox{\epsfig{figure=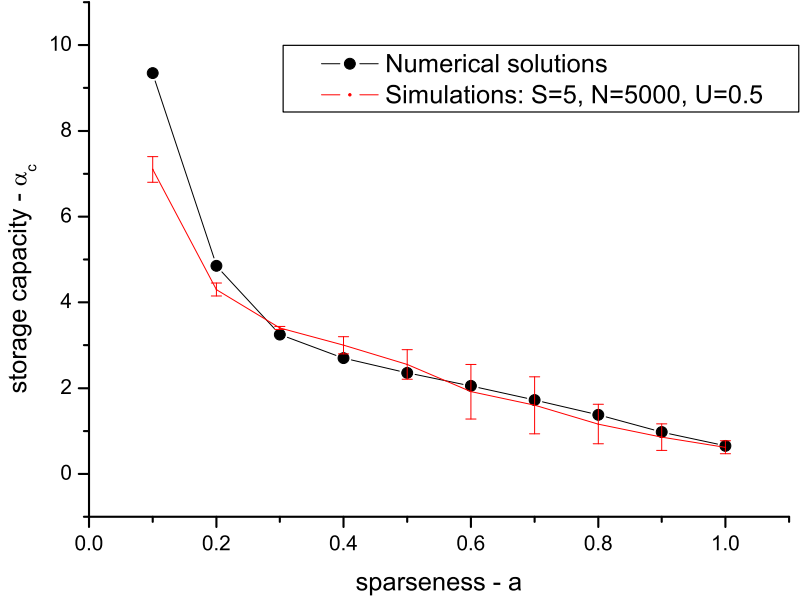,width=12cm,angle=0}}}
\caption{Dependence of the storage capacity of a sparse Potts
network of $N = 5000$ units on the sparseness $a$. The black dots
show numerical solutions of Eqs. \ref{eq:xy} and Eqs.
\ref{eq:multiple 2}, while the red line shows the result of
simulations. For very sparse simulations (low values of $a$) finite
size effects are observed, which make the storage capacity lower than
predicted by the equations.}\label{fig:sparse potts}
\end{figure}

\comment{\[
q=\frac{(1-a)}{\tilde{a}}\int Dw\int_{y\tilde{a}+x-i\sqrt{\tilde{a}}w}^{\infty}Dz\phi(z)^{S}+\int Dw\int_{-y(1-\tilde{a})+x-i\sqrt{\tilde{a}}w}^{\infty}Dz\phi(z+y)^{S}+\]
 \[
+(S-1)\int Dw\int_{y\tilde{a}+x-i\sqrt{\tilde{a}}w}^{\infty}Dz\phi(z-y)\phi(z)^{S-1}\]
 \[
m=\frac{1}{1-\tilde{a}}\int Dw\int_{-y(1-\tilde{a})+x-i\sqrt{\tilde{a}}w}^{\infty}Dz\phi(z+y)^{S}-q\frac{\tilde{a}}{1-\tilde{a}}\]
\[
C\sqrt{r}=\frac{1}{\sqrt{\tilde{\alpha}(1-\tilde{a})}}\left\{ \frac{(1-a)}{\tilde{a}}\int Dw\int_{y\tilde{a}+x-i\sqrt{\tilde{a}}w}^{\infty}Dz(z+i\sqrt{\tilde{a}}w)\phi(z)^{S}+\right.\]
 \[
+\int Dw\int_{-y(1-\tilde{a})+x-i\sqrt{\tilde{a}}w}^{\infty}Dz(z+i\sqrt{\tilde{a}}w)\phi(z+y)^{S}+\]
\begin{equation}
\left.+(S-1)\int Dw\int_{y\tilde{a}+x-i\sqrt{\tilde{a}}w}^{\infty}Dz(z+i\sqrt{\tilde{a}}w)\phi(z-y)\phi(z)^{S-1}\right\} \label{eq:multiple 2}\end{equation}}
 Putting together Eqs. \ref{eq:xy} and Eqs. \ref{eq:multiple 2}
one can construct the system of two equations that determine the
storage capacity. We show an example of their solution in Fig.
\ref{fig:sparse potts} for the parameters $U=0.5$, $S=5$ and varying
sparseness, contrasting it with simulations of a network of $N=5000$
units. This figure shows quite a good agreement between simulations
and numerical solutions for a region of the sparseness parameter
$a$, whereas for $a<0.3$ finite size effects appear, resulting in a
lower storage capacity than predicted theoretically.

\subsection{Limit case}

Given that the equations presented in the previous subsection are
quite complex, we now analyze the simpler and interesting limit case
$\tilde{a}\ll1$. Though it is not evident from the equations, the
normalized storage capacity $\tilde{\alpha}_{c}$ goes to zero in a
logarithmic way as $\tilde{a}$ goes to zero, which means that the
storage capacity is not as high as the simple signal to noise
analysis of section 3 might suggest. Our analysis of the replica
equations for the symmetric Potts model (Eq. \ref{eq:highq}) showing
logarithmic corrections is an example of this. We now analyze as
another example the sparse Potts model in the case $U=0.5$.

\begin{figure}[h]
\centerline{\hbox{\epsfig{figure=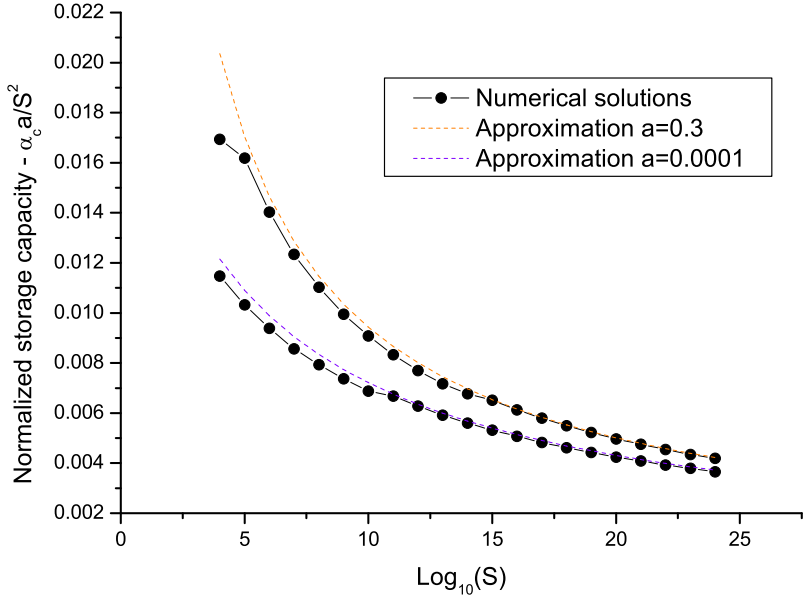,width=12cm,angle=0}}}
\caption{Corrections to the $\frac{S^{2}}{a}$ behavior of the
storage capacity of a sparse Potts network for very low values of
$\tilde{a}$ in the $U=0.5$ case. The normalized storage capacity
$\alpha_{c}a/S^{2}$ is represented, with black dots from
numerical solving Eqs. \ref{eq:xy} and Eqs. \ref{eq:multiple 2}
for two values of the sparseness: $a=0.3$ and $a=0.0001$; with color
lines from the corresponding approximation given by Eq.
\ref{eq:xreduced}.}\label{fig:high s ale}
\end{figure}

For the limit of $\tilde{a}\ll1$ one can approximate Eqs.
\ref{eq:multiple 2} by
\sublabon{equation}
\begin{eqnarray}
m&\approx&\phi(y-x)\label{eq:m}\\
q&\approx&\frac{(1-a)}{\tilde{a}}\phi(-x)+\phi(y-x)\label{eq:q}\\
C\sqrt{r}&\approx&\frac{1}{\sqrt{2\pi\tilde{\alpha}}}\left\{ \frac{(1-a)}{\tilde{a}}\exp\left(-\frac{x^{2}}{2}\right)+\exp\left(-\frac{(y-x)^{2}}{2}\right)\right\}\label{eq:c}
\end{eqnarray}
\sublaboff{equation}
which is still quite a complex system. We can now make some
self consistent assumptions. First we note that, considering $x$
and $y$ as variables that diverge logarithmically as $\tilde{a}$
goes to zero, Eqs. \ref{eq:q} and \ref{eq:c} indicate
that $\sqrt{q}\gg C\sqrt{r}$. Second, for $U=1/2$
it is possible to consider $x\approx y$, and thus, from Eq. \ref{eq:m},
$y\approx 1/\sqrt{2\tilde{\alpha}}$ and
$x\approx\varepsilon /\sqrt{2\tilde{\alpha}}$,
where $\varepsilon$ is a correcting factor for $x$ which is close
to $1$. With this in mind, and taking into account that $\tilde{\alpha}$
goes to zero with $\tilde{a}$, we can approximate Eq. \ref{eq:q} and
Eq. \ref{eq:c} by keeping only the second term in the first case and only
the first term in the second. The equations
for $y$ and $x$ can be derived from Eqs. \ref{eq:q} and Eqs. \ref{eq:xy}\[
y=\sqrt{\frac{\phi(y-x)}{\tilde{\alpha}}}\]
\begin{equation}
x=\left[2U-\frac{1-a}{\tilde{a}}\sqrt{\frac{\tilde{\alpha}}{\pi}}\exp(-\frac{x^{2}}{2})\right]\frac{1}{\sqrt{2\tilde{\alpha}}}\label{eq:xreduced}\end{equation}
Replacing $x$ by $\varepsilon/\sqrt{2\tilde{\alpha}}$ (and $\varepsilon$ by $1$ where irrelevant)we
can approximate $\tilde{\alpha}$ as\begin{equation}
\tilde{\alpha}=\frac{1}{4\:\ln\left(\frac{1}{(2U-\varepsilon)\tilde{a}}\right)}\label{eq:alfainterm}\end{equation}
Next, we posit that $\tilde{a}^{-1}$ is the larger
factor in the logarithm, while $(2U-\varepsilon)^{-1}$ gives a
correction. A rough approximation for $\alpha_{c}$
is then \begin{equation}
\alpha_{c}=\frac{S^{2}}{4\: a\ln\left(\frac{1}{\tilde{a}}\right)}\label{eq:alfa rough}\end{equation}
 which, inserted in \ref{eq:xreduced}, gives\[
(2U-\varepsilon)=(1-a)\left[4\pi\ln\left(\frac{1}{\tilde{a}}\right)\right]^{-\frac{1}{2}}\]
This expression can be re-inserted into \ref{eq:alfainterm} in order
to get a more refined approximation \begin{equation}
\alpha=\frac{S^{2}}{4\: a\ln\left(\frac{2}{\tilde{a}}\sqrt{\ln\left(\frac{1}{\tilde{a}}\right)}\right)}\label{eq:refined}\end{equation}
We show in Fig. \ref{fig:high s ale} that the approximation given by
Eq. \ref{eq:refined} fits quite well the numerical solution of
the sparse Potts model's storage capacity, particularly for very low values of
$\tilde{a}$.

\section{diluted networks}
In this section we present two modifications to our model which make
the network biologically more plausible in terms of connectivity.

First, after considering, to a zero$^{th}$ order approximation, the
long range cortical network as a {\em fully connected} network, we
now wish to describe it, to a better approximation, as a
network in which the probability that two units are connected is
$c_{M}/N$. Traditionally, analytic studies have focused on two soluble
cases: the fully connected, which we have studied in the previous
sections ($c_{M}=N$), and the highly diluted ($c_{M}\lesssim\log(N)$). A
recent work has shown, however, that the intermediate case is also
analytically treatable and that the storage capacity of an
intermediate random network, regardless the symmetry in the weights,
stands between the storage capacity of the limit cases
\cite{yasser_2004}. Supported by this result, we will focus on the (easier)
solution for the highly diluted case,
and consider any intermediate situation to be between the two limits.

The second modification reflects the notion that, although the function
of long range connections is to transmit information about the state
of a local network to another one, this transmission might not be perfectly
efficient. We thus introduce an efficacy $e$, the
probability that, in the reduced Potts model, a given state of the pre-synaptic
unit is connected with a given state of the post-synaptic
one.

Introducing these two modifications, the weights of the sparse Potts
model become\[
J_{ij}^{kl}=\frac{C_{ij}^{kl}}{c_{M}ea(1-\tilde{a})}\sum_{\mu}v_{\xi_{i}^{\mu}k}v_{\xi_{j}^{\mu}l}\]
 where $C_{ij}^{kl}$ is $1$ with probability $e\: c_{M}/N$
and $0$ otherwise.

The local field for the unit $i$ and the state $k$ can be analyzed
into a signal, a noise and a threshold part, just as in Eq. \ref{eq:field}\begin{equation}
h_{i}^{k}=\sum_{jl}J_{ij}^{kl}\delta_{\sigma_{j}l}-(1-\delta_{k0})U=(1-\delta_{k0})\left\{ (\delta_{\xi_{i}^{1}k}-\tilde{a})m_{i}^{k}+N_{k}-U\right\} \label{eq:localfield}\end{equation}
 where\[
m_{i}^{k}\equiv\frac{1}{c_{M}ea(1-\tilde{a})}\sum_{j}C_{ij}^{k\sigma_{j}}(\delta_{\xi_{j}^{1}\sigma_{j}}-\tilde{a})(1-\delta_{\sigma_{j}0})\]
Generally, when studying highly diluted networks, the noise term $N_{k}$
can be treated directly as a uniform distributed random variable,
because the states of different neurons are uncorrelated. In this
case, $N_{k}$ can not be considered as a random variable but rather
as a weighted sum of normally distributed random variables $\eta_{l}$,

\[
N_{k}\equiv\sum_{l=0}^{S}(\delta_{lk}-\tilde{a})\left\{ \sum_{\mu>1}\frac{\delta_{\xi_{i}^{\mu}l}}{c_{M}\: e(1-\tilde{a})a}\sum_{j}C_{ij}^{k\sigma_{j}}(\delta_{\xi_{j}^{\mu}\sigma_{j}}-\tilde{a})(1-\delta_{\sigma_{j}0})\right\} \equiv\sum_{l}(\delta_{lk}-\tilde{a})\eta_{l}\]
 The mean of $\eta_{l}$ is zero for all $l$ and its standard deviation
is\[
\left\langle \eta_{l}^{2}\right\rangle =\frac{N\alpha P_{l}q_{i}^{k}}{(1-\tilde{a})c_{M}\: e}\]
 with\[
q_{i}^{k}\equiv\frac{1}{c_{M}\: e\: a}\sum_{j}C_{ij}^{k\sigma_{j}}(1-\delta_{\sigma_{j}0})\]
Note that $m_{i}^{k}$ and $q_{i}^{k}$ are analogous to $m_{\rho}^{\nu}$
and $q_{\rho\lambda}$ used in Section 4. If $c_{M}\: e$ is large enough
these quantities tend to be independent of $i$ and $k$.
\begin{eqnarray}
m_{i}^{k}\rightarrow m&\equiv&\frac{1}{Na(1-\tilde{a})}\sum_{j}(\delta_{\xi_{j}^{1}\sigma_{j}}-\tilde{a})(1-\delta_{\sigma_{j}0})\nonumber\\
q_{i}^{k}\rightarrow q&\equiv&\frac{1}{N\: a}\sum_{j}(1-\delta_{\sigma_{j}0})\nonumber
\end{eqnarray}

Following the analysis of highly diluted networks in \cite{Gardner_1987},
the retrievable stable states of the network are given by the equations
\begin{eqnarray}
m&=&\frac{1}{a(1-\tilde{a})}\left\langle \left\langle \int D\mathbf{z}\sum_{\sigma}v_{\xi\sigma}\left[\frac{1}{1+\sum_{\rho\neq\sigma}\exp\left\{ \beta(h_{\rho}^{\xi}-h_{\sigma}^{\xi})\right\} }\right]\right\rangle \right\rangle \nonumber\\
q&=&\frac{1}{a}\sum_{\sigma\neq0}\left\langle \left\langle \int D\mathbf{z}\left[\frac{1}{1+\sum_{\rho\neq\sigma}\exp\left\{ \beta(h_{\rho}^{\xi}-h_{\sigma}^{\xi})\right\} }\right]\right\rangle \right\rangle \nonumber
\end{eqnarray}
where the local field, as in Eq. \ref{eq:localfield} is\[
h_{\rho}^{\xi}=m\: v_{\xi\rho}-U(1-\delta_{\rho0})+\sum_{k}\sqrt{\frac{\alpha\: N}{c_{M}\: e}\frac{q\: P_{l}}{(1-\tilde{a})}}z_{k}v_{k\rho}\]

These equations are equivalent to those obtained with the replica
method (which in the zero temperature limit are Eqs. \ref{eq:multiple fix} and
Eq. \ref{eq:HH} respectively) if one considers $C=0$ (and, thus,
$r=q$) and an effective value of $\alpha$ given by
$\alpha_{eff}=p/(c_{M}\: e)$.

\begin{figure}[h]
\centerline{\hbox{\epsfig{figure=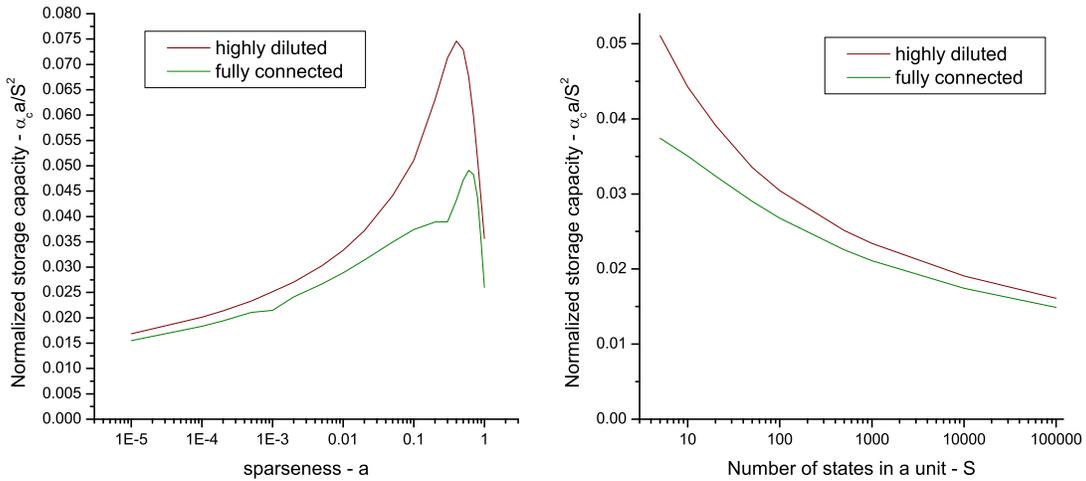,width=16cm,angle=0}}}
\caption{A comparison of the storage capacity of a fully connected
and of a highly diluted sparse Potts networks. Numerical solutions to
the corresponding equations with $U=0.5$. Left, the dependence of the
storage capacity, in the two cases, on the sparseness $a$, with $S=5$.
Right, the dependence on the number of states per
unit $S$, with $a=0.1$. In both cases we plot the normalized
storage capacity, to focus only on the corrections to
the $S^{2}/a$ behavior. Note that as $\tilde{a}\to 0$ the storage capacity
of the two types of network converges to
the same result.}\label{fig:fully vs highly}
\end{figure}

Comparing this result with that for the fully connected model one notes that, as
$\tilde{a}\to 0$, the influence of $C$ in the overall equations becomes
negligible (this can be guessed already in Eq.\ref{eq:xy} ). Therefore if
the coding is very sparse, the
fully connected and the highly diluted networks become equivalent, and
consequently also the intermediate networks. We show this in Fig.
\ref{fig:fully vs highly}. As the parameter $\tilde{a}$ goes to
zero, the storage capacity of the fully connected and the highly diluted limit
models converge.

\section{information capacity}

We have shown that the storage capacity
of well behaved models scales roughly like $S^{2}/a$, while
in the two particular examples that we analyzed in full with the replica method,
Eqs. \ref{eq:highq} and \ref{eq:alfa rough}, there is a correction
that makes it\begin{equation}
\alpha_{c}\propto\frac{S^{2}}{a\:\ln(\frac{1}{\tilde{a}})}\label{eq:scaling}\end{equation}
 for high values of $S$ and low values of $a$. We now discuss why
this is reasonable in the general case from the information storage
point of view.

It is widely believed, though not proved, that autoassociative memory
networks can store a maximum of information equivalent to a fraction
of a bit per synapse. In our model the total number of synaptic variables
is given by the different combinations of indexes of the weights $J_{ij}^{kl}$\[
number\: of\; synaptic\: variables=N\: c_{M}\: S^{2}e\]
 On the other hand, the information in a retrieved pattern is $N$
times the contribution of a single unit, which, using the distribution
in Eq. \ref{eq:distribution}, can be bounded by Shannon's entropy\[
H=-\sum_{x\in distribution}P(x)\ln\left(P(x)\right)=-\left[(1-a)\ln(1-a)+a\:\ln(\tilde{a})\right]\]
 The upper bound on the retrievable information over $p$ patterns
is then\[
information\leq-p\: N\:\left[(1-a)\ln(1-a)+a\:\ln(\tilde{a})\right]\]
 The first term between brackets is negligible with respect to the
second term provided $a$ is small enough and $S$ is large enough.
In this way we can approximate

\[
\frac{information}{number\: of\; synaptic\: variables}\leq-\frac{\alpha a\ln(\tilde{a})}{S^{2}}
\leq-\frac{\alpha_{c}a\ln(\tilde{a})}{S^{2}}\]

This result, combined with Eq. \ref{eq:scaling}, shows that the storage
capacity of our model is consistent with the idea that the information
per synaptic variable is at most a fraction of a bit.

\section{Discussion}

The capacity to store information in any device, and in particular
the capacity to store concepts in the human brain, is limited. We
have shown in a minimal model of semantic memory, and in progressive
steps, how one can expect the storage capacity to behave depending
on the parameters of the system: a global parameter - the sparseness
$a$ - and a local parameter - the number of local retrieval states
$S$, or, in other words, the storage capacity within a module.
The $S^{2}/a$ behaviour, with its corresponding logarithmic corrections,
can be thought of as the combination of two separate results: the
$a^{-1}$ behaviour due to sparseness and the $S^{2}$ behaviour of
the Potts model, which combine in a simple way. We have shown,
however, that it is not trivial to define a model that combines these
aspects correctly, and that the key is how the state
operators are defined. From this study we have deduced the minimum
requirements of any model of this kind in order to have a high capacity.
Furthermore, through the argument of information capacity we present
the well behaved family as representative of general Hebbian models
with the same degree of complexity.

The featural representation approach has been so far successful in 
explaining several phenomena associated to semantic memory, like 
similarity priming, feature verification, categorization and conceptual 
combination \cite{McRae_2005, McRae_1997}. The present work demonstrates that the 
advantage of the use of features in allowing the representation of
a large number of concepts can be realized in a simple associative memory network. 
More quantitatively, our calculation specifies that in the Potts model the 
number of concepts that can be stored is neither linear \cite{OKa+92b} 
nor an arbitrary power \cite{Schyns_1997} of the number $S$ of values a feature can take,
but quadratic in $S$.

In the case of non-unitary sparseness, one can associate the necessity
of introducing a threshold ($U$) term, whatever its exact form in
the local field or the Hamiltonian, with a criterion of selectivity,
which is actually observed in the representation of concepts in the
brain, as pointed out in the introduction. The threshold behaviour,
which is a typical characteristic of neurons, appears to be also necessary
at the level of local networks in order to maintain activity low in
the less representative modules. The origin of such a threshold has
not been discussed in this paper. However, a comment on this issue
can be made regarding the internal dynamics of local networks. One
can show that, as extensively described in the literature \cite{Amit_1989},
only when the state of a local autoassociative network is driven by
external fields sufficiently close to an attractor (inside one of
the $S$ basins of attraction) the local system may end up retrieving
a pattern on its own, a process that from the global network
point of view corresponds to the activation of a unit. The local basin boundary
acts in the full system as an effective threshold, roughly equivalent to the 
simple $U$ term we introduced in the local field of our reduced system.
Whether this threshold mechanism is enough, or some addition must
be made, can be assessed by studying, in the future, the complete
multimodular network without reducing it to Potts units.

\begin{acknowledgments}
This research
was supported by Human Frontier Grant RGP0047/2004-C. 
We thank  Yasser Roudi and David Hansel for very useful conversations. 
\end{acknowledgments}
\bibliographystyle{revtex}
\bibliography{emilioj}

\end{document}